\documentclass[conference]{IEEEtran}
\usepackage{cite}
\usepackage{amsmath,amssymb,amsfonts}
\usepackage{algorithmic}
\usepackage{graphicx,color}
\usepackage{textcomp}
\usepackage{xcolor}
\usepackage{hyperref}
\usepackage{algorithm,algorithmic}
\usepackage{tikz}
\usetikzlibrary{arrows.meta, positioning}
\usepackage{multirow}
\usepackage{siunitx}
\usepackage{booktabs}
\usepackage{tabularx}
\usepackage{makecell}
\usepackage{fix-cm}
\usepackage[printonlyused]{acronym}

\def\BibTeX{{\rm B\kern-.05em{\sc i\kern-.025em b}\kern-.08em
    T\kern-.1667em\lower.7ex\hbox{E}\kern-.125emX}}
\AtBeginDocument{\definecolor{tmlcncolor}{cmyk}{0.93,0.59,0.15,0.02}\definecolor{NavyBlue}{RGB}{0,86,125}}

\begin{document}


\title{Toward a Stable and Deployable Adaptive Chirplet Transform: Residual Projection, Hybrid GPU Acceleration, and Multi-Channel Scalability}

\author{\IEEEauthorblockN{Nishant Kumar}
\IEEEauthorblockA{\textit{University of Toronto} \\
Toronto, Canada \\
0009-0007-7169-9241}
\and
\IEEEauthorblockN{Steve Mann}
\IEEEauthorblockA{\textit{University of Toronto} \\
Toronto, Canada \\
0000-0003-0363-3690}
}


\maketitle

\begin{abstract}
The Adaptive Chirplet Transform is a flexible framework that can decompose non-stationary signals into sparse chirplets; it has been applied to signals such as electroencephalography, electromyography and radar. However, the practical deployment of this transform has been hindered by two challenges: algorithmic instability in prior implementations, which can lead to divergent decompositions, and the computational cost of searching over a high-dimensional parameter space. This paper addresses both by a sequence of contributions. Firstly, unit normalization and residual-based projection are introduced to align the decomposition with Matching Pursuit Theory, thereby eliminating divergence and substantially reducing residual error across all signal domains, as demonstrated on three representative signal types. A hybrid CPU--GPU architecture offloads chirplet family generation to the CPU while parallelizing the search on the GPU, removing bottlenecks in CPU-only search and GPU-only generation, achieving speedups of $\mathbf{6.6-7.38\times}$ on desktop hardware, with consistent gains observed across laptop and embedded platforms. Multichannel batching enabled simultaneous multi-signal processing, amplifying the speedup, which scaled from $\mathbf{3.94\times}$ for a single channel to $\mathbf{8.22\times}$ at 10 channels.  Finally, a hierarchical coarse-to-fine search, inspired by Logon Expectation Maximization, is introduced. This reduced peak memory usage below 1 GB while maintaining similar reconstruction quality, at the cost of longer runtime. Together, these contributions establish a correct, stable and practically deployable foundation for chirplet-based signal decomposition.
\end{abstract}

\section{Introduction}

\IEEEPARstart{N}{on-stationary} signals are ubiquitous across biomedical, radar, and industrial sensing domains. The current methods of analyzing these signals, such as the short-time Fourier Transform or the wavelet transform, impose a fixed time-frequency resolution. This limits their ability to adapt to drifting signal structures. The \ac{ACT} is grounded in the theory of the chirplet transform that was introduced by Mann and Haykin~\cite{mannVI91, mannsp}. The transform decomposed signals into a sparse set of chirplets, each parameterized by time centre, frequency centre, chirp rate, and log-duration, thereby enabling adaptation to time-varying frequency content. This flexibility has motivated applications across \ac{EEG} analysis~\cite{1643406, JIANG2021102699}, mechanical fault detection~\cite{MA2021109574}, and radar~\cite{4606910}, among others.

However, the \ac{ACT} remains computationally demanding, as the search scales with the product of all grid sizes. This makes high-resolution decomposition impractical for CPU-based processing of long signals or large channel counts~\cite{9398775}. Previous work~\cite{11512436} investigated a GPU-only pipeline for the \ac{ACT}, achieving up to $18\times$ speedups in the search relative to CPU across a range of simulated signal types. In this work, a bottleneck was discovered: chirplet generation in a pure GPU pipeline took significantly longer than on the CPU, likely due to kernel launch overhead. Furthermore, the prior implementation projected the chirplets onto the original signal rather than onto the current residual and did not apply unit normalization, leaving convergence uncertain and departing from the theoretical requirements of \ac{MP}.

This paper addresses both of these limitations of the prior implementation and further extends the framework through a sequence of additional contributions.\footnote{Portions of the residue-based \ac{ACT} formulation and preliminary hybrid CPU--GPU results presented in this work were first developed in the author's MASc thesis~\cite{kumar2026thesis}.} First, unit normalization and residual-based projection are introduced to align the \ac{ACT} with standard \ac{MP} theory. This established a stable decomposition, which was the foundation for reliable acceleration. Then, a hybrid CPU--GPU architecture was proposed to leverage both CPU and GPU by assigning chirplet generation and the projection search to their respective processors, thereby retaining the search speedup of the prior implementation. With reliable acceleration, multichannel batching was introduced to enable simultaneous multi-signal processing, with GPU speedup amplifying as the channel count grows. This can enable practical, real-time deployment in multi-channel settings, such as clinical \ac{EEG}. Additionally, hardware scalability is evaluated by comparing high-end desktop consumer GPUs with consumer laptops and a 2018 embedded platform, demonstrating that the framework is not limited to high-end laboratory hardware. The final contribution was a hierarchical coarse-to-fine search, inspired by \ac{LEM}, serving as a preliminary exploratory decomposition that maintained competitive reconstruction accuracy compared with the original algorithm while reducing peak memory usage to below 1 GB. All contributions were validated on publicly available datasets.

The main contributions of this paper are summarized as follows:

\begin{itemize}
    \item Unit normalization and residual projection, which resolved divergent decomposition behaviour in prior implementations, establishing a stable foundation for acceleration.
    \item A hybrid CPU--GPU architecture resolving the bottleneck of generating the chirplets on the previous GPU-only approach, while achieving $\times 7.38$ speedup per epoch on a desktop RTX 5070.
    \item Multichannel batched signal processing with GPU speedup amplifying to $\times 8.22$ at 10 channels from $\times 3.94$.
    \item Hardware scalability evaluation across desktop, laptop and embedded platforms, confirming the practicality of deploying this algorithm outside of laboratory environments.
    \item A tunable LEM-inspired hierarchical coarse-to-fine search which reduced peak memory below 1 GB, with tradeoffs between memory, speed and reconstruction quality.
\end{itemize}

The remainder of this paper is organized as follows. Section~\ref{sect:background} provides
background on the chirplet transform, \ac{MP}, and GPU computing. Section~\ref{sect:Algorithm} introduces the residue-based \ac{ACT} algorithm and validates it across three signal domains. Section~\ref{sect:accel} presents the hybrid CPU--GPU acceleration framework and benchmarks. The following sections address multichannel scalability, hardware portability, and memory-efficient search, respectively. Section~\ref{sect:Discussion} presents an overall discussion, and Section~\ref{sect:Conclusion} concludes the paper.

\subsection{ABBREVIATIONS AND ACRONYMS}\vspace{0.2em}
\begin{acronym}[ACT] 
    \acro{ACT}{Adaptive Chirplet Transform}
    \acro{MP}{Matching Pursuit}
    \acro{EEG}{Electroencephalography}
    \acro{EMG}{Electromyography}
    \acro{CUDA}{Compute Unified Device Architecture}
    \acro{OpenCL}{Open Computing Language}
    \acro{LHS}{Latin Hypercube Sampling}
    \acro{LEM}{Logon Expectation Maximization}
    \acro{BFGS}{Broyden–Fletcher–Goldfarb–Shanno}
\end{acronym}

\section{Background}
\label{sect:background}
\subsection{Chirplet Transform}

\begin{figure}[!ht]
    \centering
    \includegraphics[width=\linewidth]{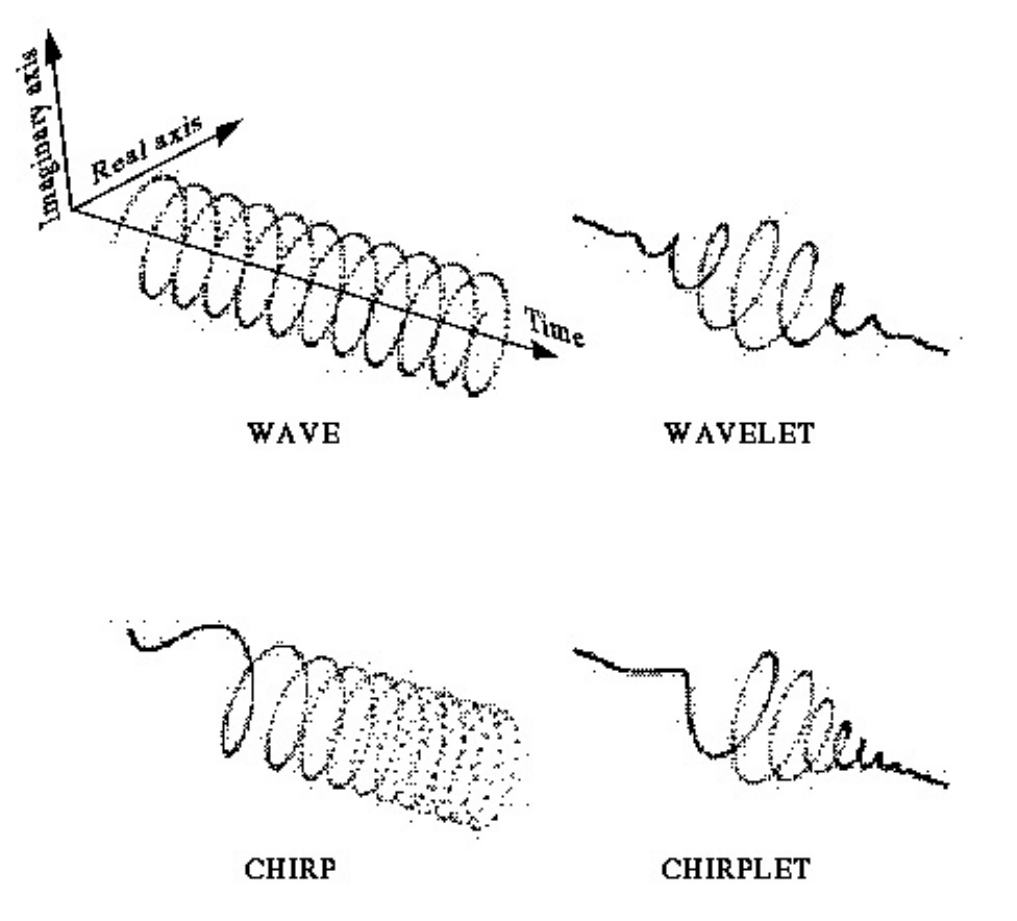}
    \caption{Conceptual relationship between waves, wavelets, chirps, and chirplets. A wave has constant frequency; a wavelet localizes oscillations in time; a chirp introduces frequency variation; and a chirplet combines both localization and frequency modulation.  Figure reproduced from~\cite{mannsp}.}
    \label{fig:conceptual}
\end{figure}

The chirplet transform, introduced in~\cite{mannVI91}, generalizes the Gabor and wavelet transforms to accommodate signals with time-varying spectral content (see Figure~\ref{fig:conceptual}). In its most general form, chirplets are based on five parameters: a time center $t_c$, a frequency center $f_c$, a chirp rate $c$, a dispersiveness $d$, and a log-duration $\log\Delta_t$. The chirp rate describes the modulation of frequency over time, corresponding to a shear along the frequency axis of the time-frequency plane. Dispersiveness describes a frequency-varying time shift, corresponding to a shear along the time axis on that same plane. Together, these five parameters span the metaplectomorphisms of the time-frequency plane, covering all symplectic transformations of the time-frequency plane while preserving the underlying structure of time-frequency analysis~\cite{mannsp}.

In practice, the chirplet is usually specialized to a Gaussian envelope, which can be represented by an ellipse on the time-frequency plane, whose area saturates the bound imposed by the Heisenberg uncertainty principle. The ellipse can be rotated, and its aspect ratio varied, which geometrically represents chirp rate and dispersiveness, respectively, but the two cannot act independently on the Gaussian envelope. This coupling reduces the degrees of freedom from five to four. A family of Gaussian chirplets is therefore parametrized by $(t_c, f_c, c, \log\Delta_t)$, giving:

\begin{equation}
\label{eq1}
\resizebox{0.95\linewidth}{!}{$
g_{t_c,\, f_c,\, \log\Delta_t,\, c}(t)
=
\frac{1}{\sqrt{\smash[b]{\sqrt{\pi}}\;\Delta_t}}
e^{-\frac{1}{2}\left(\frac{t-t_c}{\Delta_t}\right)^2}
e^{\,i2\pi\left[c(t-t_c)^2+f_c(t-t_c)\right]}
$}
\end{equation}

Any signal can then be approximated as a weighted superposition of such chirplets,

\begin{equation}
x(t) \approx \sum_{n=1}^{N} \alpha_n \, g(t;\, t_{c,n},\, f_{c,n},\, c_n,\, 
\log\Delta_{t,n})
\end{equation}

where $\alpha_n$ are the decomposition coefficients. Each chirplet is matched to the local time-frequency structure of the signal. The chirplet transform has demonstrated utility across domains including EEG analysis~\cite{1643406, JIANG2021102699}, mechanical 
fault detection~\cite{MA2021109574}, and radar~\cite{4606910}. However, the computational cost of searching a high-resolution parameter space remains a practical barrier~\cite{9398775}, motivating the framework presented in this work.

\subsection{Matching Pursuit}
\ac{MP} was introduced by Mallat and Zhang~\cite{258082}. It provides a greedy iterative algorithm for decomposing a signal as a linear combination of basis functions from an overcomplete dictionary $\mathcal{G}$. At each iteration, it searches for the basis function that maximizes the inner product with the current residual, then computes the coefficient that represents the contribution of the selected basis function to the residual. The projection is then subtracted from the residual before the next iteration. \ac{MP} and its extensions have become foundational tools widely used to represent signals in applications such as image processing, biomedical signal analysis, and audio coding ~\cite{529037, 9100290, 7080374}.

\subsection{General Purpose Graphics Processing Units}
General-purpose computing on graphics processing units (GPGPU) uses the parallel architecture of modern GPUs to accelerate compute-intensive workloads beyond conventional graphics rendering. GPUs consist of thousands of lightweight cores to execute many threads concurrently, whereas CPUs were optimized for low-latency execution of sequential tasks. So for data-parallel problems where the same operation is being applied, GPUs provide a substantial advantage over CPUs. Frameworks such as \ac{CUDA} and \ac{OpenCL} enable this parallelism for general-purpose workloads, where computations are expressed as kernels that run across large collections of threads~\cite{6419068}. Achieving the highest performance requires very careful consideration of GPU memory, specifically registers, shared memory and minimizing data transfer between the host and the device. 

The chirplet-based matching pursuit framework exhibits substantial inherent parallelism, particularly in evaluating inner products between the signal residual and candidate chirplets, which can be computed independently for each parameter combination. This has been shown to map naturally onto GPU architectures, enabling concurrent processing of the chirplet set. In this work, the implementation was developed in Python using the CuPy library, which provides optimized, NumPy-compatible arrays for GPU operations. This library allows for efficient development without requiring explicit low-level kernel programming. Even though it may provide much less direct control than \ac{CUDA} implementations, it offers a balance between performance and implementation complexity.

\section{Residue-Based ACT Algorithm}
\label{sect:Algorithm}
\subsection{Algorithm}

As part of the \ac{ACT}, a signal $s(t)$ is decomposed to a weighted sum of chirplet components called ``chirplets''~\cite{mannVI91},

\begin{equation}
    s(t) \approx \sum_{k=1}^{P} c_k \, g(\theta_k^*),
\end{equation}

where each component $g(\theta)$ is part of a parametric family of chirplets $\mathcal{G} = \{g(\theta)\}$, such that $\theta$ represents $(t_c, f_c, c, \log\Delta_{t})$.  This family of chirplets is continuous over the parameter space. In practice, however, there is a finite discrete subset of the family referred to as a dictionary, as seen in many \ac{MP} frameworks  ~\cite{258082}. This subset is constructed by sampling the parameter space within user-specified ranges and serves as the search space for each iteration.

In prior \ac{ACT} work ~\cite{9398775} and its contemporaneous implementation ~\cite{aman_chirplet}, this set was precomputed, and a coarse search then identified the chirplet that best represented the signal via the inner product with the original signal $s(t)$. This was then refined using \ac{BFGS}, and the corresponding coefficient was then computed as $c = \langle g^*, s(t) \rangle$. This process repeats for each $ P$-decomposition order. However, in the implementation, the chirplet is projected onto the original signal and lacks a normalization constraint, departing from the theoretical requirements of MP and potentially leaving convergence uncertain.

Building on theoretical \ac{MP} work, the residue-based \ac{ACT} is introduced to address these limitations. As such, all chirplets are normalized to unit $\ell_2$ norm prior to search. Normalization is critical in this process, because without it, it would favour atoms with larger $\ell_2$ norm regardless of how well they match the shape of the residual. At each order $k$, the selection and projection are performed on the current residual rather than the original signal, except for the first order, where $r_0 = s(t)$, and

\begin{equation}
    r_{k+1} = r_k - c_k \, g^*, \qquad 
    a_{k+1} = a_k + c_k \, g^*.
\end{equation}

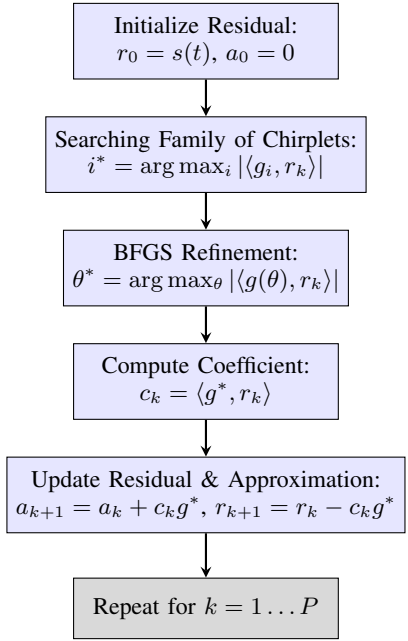
\begin{figure}[!ht]
\centering
\begin{tikzpicture}[node distance=1.5cm, auto, font=\small, >=stealth]
\tikzstyle{block} = [rectangle, draw, fill=blue!10, minimum width=3.5cm, 
    minimum height=1cm, align=center]
\tikzstyle{subblock} = [rectangle, draw, fill=gray!30, minimum width=3.5cm, 
    minimum height=0.8cm, align=center]
\tikzstyle{arrow} = [->, thick]
\node [block] (residual) {Initialize Residual:\\ $r_0 = s(t)$, $a_0 = 0$};
\node [block, below of=residual] (search) {Searching Family of Chirplets:\\ 
    $i^* = \arg\max_i |\langle g_i, r_k \rangle|$};
\node [block, below of=search] (minimize) {BFGS Refinement:\\ 
    $\theta^* = \arg\max_\theta |\langle g(\theta), r_k \rangle|$};
\node [block, below of=minimize] (coeff) {Compute Coefficient:\\ 
    $c_k = \langle g^*, r_k \rangle$};
\node [block, below of=coeff] (update) {Update Residual \& Approximation:\\ 
    $a_{k+1} = a_k + c_k g^*$, $r_{k+1} = r_k - c_k g^*$};
\node [subblock, below of=update] (repeat) {Repeat for $k = 1 \dots P$};
\draw [arrow] (residual) -- (search);
\draw [arrow] (search) -- (minimize);
\draw [arrow] (minimize) -- (coeff);
\draw [arrow] (coeff) -- (update);
\draw [arrow] (update) -- (repeat);
\end{tikzpicture}
\caption{Simplified residue-driven ACT iteration showing the main computational steps of searching the family of chirplets.}
\label{fig:act_simple_flow}
\end{figure}

The coarse search over the chirplet family first selects the best candidate $i^* = \arg\max_i |\langle g_i, r_k \rangle|$, which then initializes the refinement through \ac{BFGS} solving $\theta^* = \arg\max_\theta |\langle g(\theta), r_k \rangle|$ against the current residual signal. This procedure is summarized in Fig.~\ref{fig:act_simple_flow}, and the changes between the previous and current implementations are summarized in Table~\ref{tab:act_comparison}.

\subsection{Methodology}
Experiments were conducted on three publicly available datasets with different physiological domains: \ac{EEG}, \ac{EMG}, and radar micro-Doppler signatures, to demonstrate that this residue-based \ac{ACT} can operate as a decomposition framework across heterogeneous domains, consistent with \ac{MP} principles.

\begin{table}[!ht]
    \centering
    \renewcommand{\arraystretch}{1.4}
    \resizebox{\linewidth}{!}{
    \begin{tabular}{|l|c|c|}
    \hline
    \textbf{Property} & \textbf{Previous ACT} & \textbf{Residue-Based ACT} \\
    \hline
    Atom normalization & None & Unit L2 norm \\
    \hline
    Projection target & Original signal & Current residual \\
    \hline
    Monotone residual decrease  & Not enforced & Aligned with MP theoretical conditions \\
    \hline
    Caching Available & Yes & Yes \\
    \hline
    BFGS refinement & Yes & Yes \\
    \hline
    \end{tabular}
    }
    \vspace{1mm}
    \caption{Comparison of algorithmic properties between the previous ACT implementation~\cite{9398775} and the residue-based ACT introduced in this work.}
    \label{tab:act_comparison}
\end{table}

\subsubsection{Datasets}
\ac{EEG} data were collected from the Bitbrain Wearable EEG headset during sleep using the first patient's recording~\cite{ds005555:1.1.0}. \ac{EMG} data was taken from the NinaPro DB2 dataset~\cite{Atzori_Gijsberts_Castellini_Caputo_Hager_Elsig_Giatsidis_Bassetto_Müller_2014}, specifically the basic finger movement data from Patient 1. The radar data were obtained from the micro-Doppler signature of a DJI Matrice 200 (D1) ~\cite{41e8-8v73-25}, specifically the peak range cell, which is expected to have the highest signal-to-noise ratio. 

\subsubsection{Preprocessing}
Each signal was first preprocessed using zero-phase filtering to avoid phase distortion, which is especially important because phase inconsistencies can affect chirplet-based analysis. \ac{EEG} data was processed using MNE-Python's notch filter at 50 Hz to remove the European power-line frequency, which was then followed by a bandpass filter between 0.1 Hz and 20 Hz to retain the majority of the physiologically relevant content within the delta to beta frequency bands. \ac{EMG} data was first notch-filtered at the 9 harmonics of 50 Hz using a second-order IIR filter with a quality factor of 30, followed by a fourth-order Butterworth bandpass filter to capture most neuromuscular activation while removing low-frequency motion artifacts. The last harmonic (450 Hz) coincides with the upper cutoff of the bandpass, which means that both filters attenuated the signal. For both \ac{EEG} and \ac{EMG}, the first channel was kept for analysis. Radar data was bandpass filtered using a fourth-order Butterworth filter between 10 Hz and the Nyquist frequency of approximately 8533 Hz. These filters were chosen to provide reasonable signal conditioning while preserving the signal's underlying structure.

\subsubsection{Parameter Selection}
The chirplet parameter ranges that were used are summarized in Table~\ref{tab:parameters_clean}. These were selected as principled, approximate estimates that align with the signal's characteristics rather than as optimized values. The window lengths were primarily chosen to maintain consistency between all three domains. These listed ranges are intended as a reproducible baseline that practitioners with domain expertise can further refine to more accurately reflect the temporal and spectral characteristics of their specific application. The time-centre parameter spanned the full analysis window in all cases, ensuring that chirplets could be localized at any position. The frequency-centre parameter was chosen to reflect the dominant signal content based on the pre-processing. The log-duration parameter was kept constant to correspond to narrow transient bursts to slow-frequency-drifting oscillations, and the chirp-rate symmetric parameter range was set to capture both upward and downward frequency sweeps. 

These ranges, shared across all three domains, serve as a reasonable default, without considering domain-specific behaviours on the time-frequency plane. The chosen step sizes can reflect a trade-off between search granularity and computational cost. If the steps are coarser, as seen in the 100 Hz for Radar, it would reduce the dictionary size and runtime but may miss the optimal chirplet; this can be partially mitigated by the \ac{BFGS} step, which can recover sub-grid precision through continuous optimization in the neighbourhood of the coarse candidate.

\subsubsection{Experimental Protocol}
For each signal, the first 50 non-overlapping consecutive windows were taken. The lengths of the windows reflected the range of time centres in Table~\ref{tab:parameters_clean}, so for EEG it was set to 256 samples, which is considered 1 second of data; for EMG it was set to 200 samples, which was $100~ms$; and for radar it was set to 256 samples, which is considered $15~ms$. These windows ensure that each chirplet atom is well matched, enabling evaluation of chirplet reconstruction fidelity. This set of 50 windows was evaluated 5 times, yielding 250 evaluations per signal, enabling runtime assessment through repeated sets and verification that accuracy remained stable across multiple epoch windows. Decomposition was performed at both the 5th and 10th orders to assess whether additional components could yield further reconstruction, using normalized residue
\begin{equation}
\mathcal{E}_k = \frac{|r_k|^2}{|s|^2},
\end{equation}
where a value approaching zero would signify better reconstruction. The results are compared with the prior ACT implementation ~\cite{aman_chirplet} to quantify the impact of residual-based projection and unit normalization. Furthermore, to test convergence behaviour, a single epoch from each of the three signals was decomposed iteratively from 1 to 200 in steps of 2, and the resulting normalized residue was plotted as a function of order to identify any visible trends and the rate of energy capture. Both experiments were run on an Apple M1 Max Mac Studio with 32 GB of memory. \footnote{All code and experimental scripts are publicly available  at \url{https://github.com/nishantkumar201/adaptive-chirplet-transform}}

\begin{table}[!ht]
\centering
\small
\resizebox{\linewidth}{!}{
\begin{tabular}{lcccc}
\toprule
\textbf{Signal} & \textbf{Parameter} & \textbf{Range / Values} & \textbf{Step} & \textbf{\makecell{Parameter \\Combinations}}\\
\midrule
EEG & \texttt{tc\_info} & 0 -- 256 & 16 & \multirow{4}{*}{185,600}\\
EEG & \texttt{fc\_info} & 0.5 -- 15 & 0.5 \\
EEG & \texttt{logDt\_info} & -4 -- 1 & 0.5 \\
EEG & \texttt{c\_info} & -10 -- 10 & 0.5 \\
\midrule
EMG & \texttt{tc\_info} & 0 -- 200 & 32 & \multirow{4}{*}{113,778}\\
EMG & \texttt{fc\_info} & 20 -- 450 & 10 & \\
EMG & \texttt{logDt\_info} & -4 -- 0 & 0.3 \\
EMG & \texttt{c\_info} & -10 -- 10 & 0.75 \\
\midrule
Radar & \texttt{tc\_info} & 0 -- 256 & 16 & \multirow{4}{*}{520,128}\\
Radar & \texttt{fc\_info} & 10 -- 8533 & 100 \\
Radar & \texttt{logDt\_info} & -4 -- 0 & 0.3 \\
Radar & \texttt{c\_info} & -10 -- 10 & 0.75 \\
\bottomrule
\end{tabular}
}
\vspace{1mm}
\caption{Parameter ranges and step sizes for EEG, EMG, and Radar signals used in the ACT experiments.}
\label{tab:parameters_clean}
\end{table}

\subsection{Results and Discussion}
\begin{figure}[!ht]
\centering
\includegraphics[width=\linewidth]{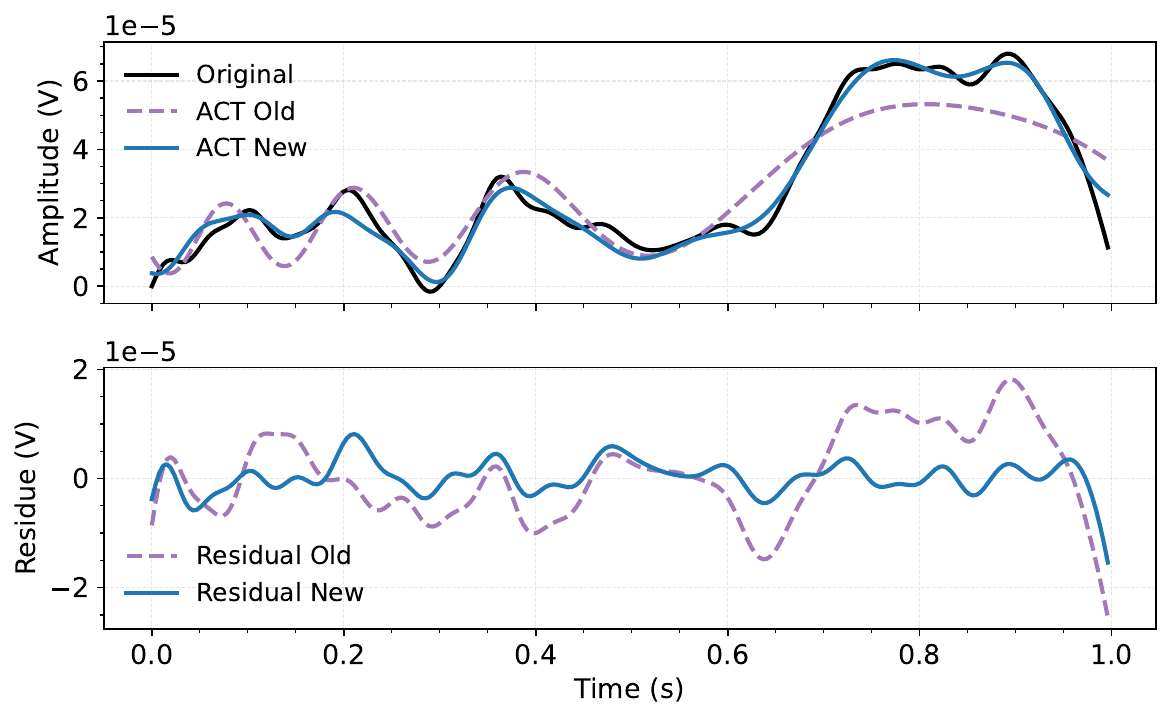}
\caption{Representative EEG epoch showing signal approximation and residuals. Black: original signal. Purple dashed: old ACT approximation. Blue: residue-driven ACT approximation. The residue-driven ACT more closely tracks rapid signal variations, resulting in lower residual amplitude.}
\label{fig: EEG+approx}
\end{figure}

Table~\ref{tab:performance_summary_clean} summarizes results of normalized residue and runtimes across all 3 signals with the decomposition orders of 5 and 10. The newer ACT consistently had lower normalized residue than the prior implementation across all test cases, suggesting that the new method yields a more reliable decomposition. One interesting test case that highlights the contrast between the two methods is the EEG results at order 10, where the previous implementation yields a normalized residue of $1.417 \pm 1.017$. The large mean and variance indicate instability in the decomposition across epochs. This behaviour is consistent with a lack of unit normalization, where if the selected chirplet carries a large $\ell_2$ norm, the update term can overshoot the residual energy, especially when adding more decomposition orders. This can cause $\|r_k\|^2 > \|s\|^2$ and a normalized residue exceeding unity. As seen in the normalized residue value, this occurs occasionally, suggesting it depends on the characteristics of the epoch's signal. The newer implementation eliminates this instability, which reduces the \ac{EEG} normalized residue to $0.031 \pm 0.012$. This improvement is further demonstrated in Figure ~\ref{fig: EEG+approx}, which shows that the residual-driven approximation more closely tracks the EEG epoch at a decomposition order of 5 than the original implementation, resulting in a lower residual amplitude.

\begin{table*}[t]
\centering
\small
\begin{tabularx}{\textwidth}{cc>{\centering\arraybackslash}X>{\centering\arraybackslash}X>{\centering\arraybackslash}X>{\centering\arraybackslash}X}
\toprule
\textbf{Signal} & \textbf{Order} & 
\textbf{Old Norm. Residue} & \textbf{New Norm. Residue} & 
\textbf{Old Full Search Runtime (s)} & \textbf{New Full Search Runtime (s)} \\
\midrule
EEG   & 5  & 0.461 $\pm$ 0.335 & 0.076 $\pm$ 0.029 & 8.16 $\pm$ 0.12  & 4.75 $\pm$ 0.07 \\
EEG   & 10 & 1.417 $\pm$ 1.017 & 0.031 $\pm$ 0.012 & 13.11 $\pm$ 0.49 & 5.53 $\pm$ 0.09 \\
EMG   & 5  & 0.961 $\pm$ 0.012 & 0.618 $\pm$ 0.053 & 15.58 $\pm$ 0.65 & 8.32 $\pm$ 0.11 \\
EMG   & 10 & 0.928 $\pm$ 0.019 & 0.448 $\pm$ 0.045 & 29.83 $\pm$ 0.84 & 13.67 $\pm$ 0.43 \\
Radar & 5  & 0.972 $\pm$ 0.003 & 0.179 $\pm$ 0.063 & 18.14 $\pm$ 0.51 & 17.46 $\pm$ 0.14 \\
Radar & 10 & 0.945 $\pm$ 0.005 & 0.144 $\pm$ 0.055 & 28.16 $\pm$ 2.55 & 24.70 $\pm$ 0.23 \\
\bottomrule
\end{tabularx}
\caption{Performance summary for EEG, EMG, and Radar signals at orders 5 and 10.
Normalized residue reductions and full runtime are shown
for the new method compared to the old method.}
\label{tab:performance_summary_clean}
\end{table*}

Looking at the runtime in the table, the newer implementation showed consistently faster epoch runtimes for \ac{EEG} and \ac{EMG}, while improvements for Radar were minimal. This was an unintended consequence of modifying a type cast in the prior code, which involved a redundant conversion between double- and single-precision. The minimal runtime improvement observed for radar is due to the coarse search dominating the compute budget compared to the cast correction, which has 520,128 parameter combinations (Table~\ref{tab:parameters_clean}). Interestingly, the runtime for \ac{EMG} is longer than that of \ac{EEG}, despite a comparable number of combinations. The source of this overhead is less clear; it may relate to the broader frequency range, which can produce a more complex \ac{BFGS} cost landscape, but further profiling is required. 

\begin{figure}[!ht]
    \centering
    \includegraphics[width=\linewidth]{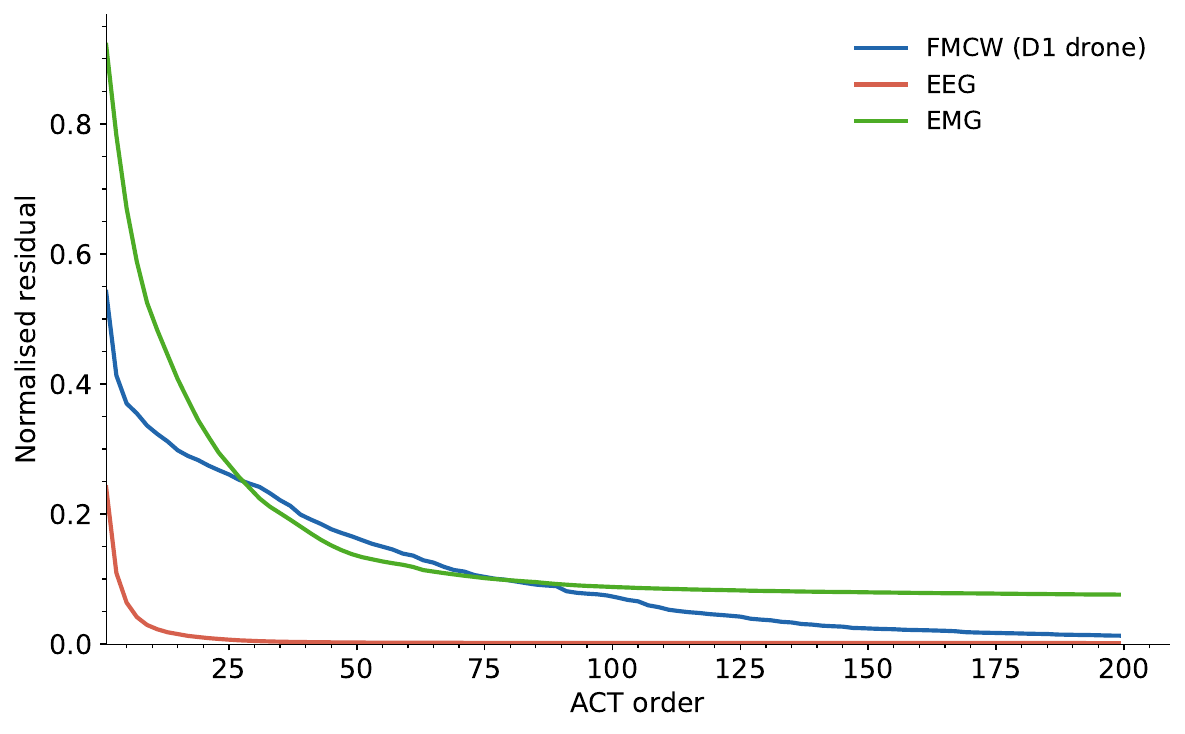}
    \caption{Normalized residual energy as a function of ACT decomposition order for three signal domains: EEG (HB$_1$), FMCW radar (DJI Matrice 200), and surface EMG (basic movement). The normalized residual is the ratio of the residual to the total signal energy after P-order decomposition.}
    \label{fig:ordervsresidue}
\end{figure}

Convergence behaviour for all three domains is illustrated in Figure ~\ref{fig:ordervsresidue}. Looking at \ac{EEG}, it decreases rapidly and approaches 0 around order 50, indicating that the chirplet atoms and the chosen ranges are a good match for this signal. The \ac{EMG} exhibits a different behaviour: the normalized residue decreases but then plateaus at around 75 orders at 0.1, indicating it captures roughly 90\% of the signal. This could be due to the selected chirplet range not capturing the entire signal in this epoch, suggesting that the \ac{EMG} requires higher chirpiness or finer resolution. Radar converges more gradually, but it reaches near 0 at around order 200. The rate of convergence could be affected by the coarse 100 Hz frequency steps, which can limit dictionary precision, but this is still partially compensated for by the \ac{BFGS} stage. Now that reliable decomposition has been demonstrated across domains, further acceleration is required for practical deployment; hence, the next section implements it using GPU parallelism.

\section{Accelerated Framework}
\label{sect:accel}
\subsection{Algorithm}
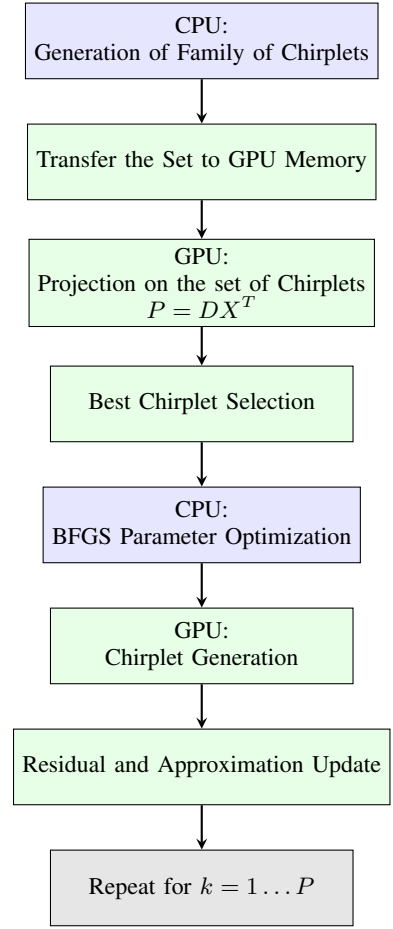
\begin{figure}[!ht]
\centering
\begin{tikzpicture}[node distance=1.6cm, auto, font=\small, >=stealth]
\tikzstyle{cpu} = [rectangle, draw, fill=blue!10, minimum width=4cm, 
    minimum height=1cm, align=center]
\tikzstyle{gpu} = [rectangle, draw, fill=green!10, minimum width=4cm, 
    minimum height=1cm, align=center]
\tikzstyle{iter} = [rectangle, draw, fill=gray!20, minimum width=4cm, 
    minimum height=1cm, align=center]
\tikzstyle{arrow} = [->, thick]
\node[cpu] (dict) {CPU:\\ Generation of Family of Chirplets};
\node[gpu, below of=dict] (transfer) {Transfer the Set to GPU Memory};
\node[gpu, below of=transfer] (projection) {GPU:\\ Projection on the set of Chirplets\\ 
    $P = D X^T$};
\node[gpu, below of=projection] (selection) {Best Chirplet Selection};
\node[cpu, below of=selection] (bfgs) {CPU:\\ BFGS Parameter Optimization};
\node[gpu, below of=bfgs] (atom) {GPU:\\ Chirplet Generation};
\node[gpu, below of=atom] (update) {Residual and Approximation Update};
\node[iter, below of=update] (repeat) {Repeat for $k = 1 \dots P$};
\draw[arrow] (dict) -- (transfer);
\draw[arrow] (transfer) -- (projection);
\draw[arrow] (projection) -- (selection);
\draw[arrow] (selection) -- (bfgs);
\draw[arrow] (bfgs) -- (atom);
\draw[arrow] (atom) -- (update);
\draw[arrow] (update) -- (repeat);
\end{tikzpicture}
\caption{Hybrid CPU--GPU execution pipeline for the accelerated ACT. The generation and parameter refinement are performed on the CPU, while the projection and residual updates are GPU-accelerated.}
\label{fig:act_hybrid_pipeline}
\end{figure}

Not all stages of the ACT benefit equally from GPU execution. Based on prior work~\cite{11512436}, chirplet generation was found to be faster on the CPU, whereas chirplet family search seemed to benefit from the GPU, motivating the hybrid architecture proposed here. This distinction likely arises because generating each chirplet requires independent operations that map poorly to GPU architectures, incurring kernel launch overhead. However, the coarse search can be viewed as a large matrix projection $P = DX^T$, with $D \in \mathbb{R}^{K \times L}$ and $X \in \mathbb{R}^{B \times L}$, which maps naturally to GPU parallelism. Therefore, the proposed hybrid architecture assigns chirplet generation to the CPU using NumPy, and once the chirplet family has been created, it transfers them to the GPU's VRAM for caching. The projection and residual update stages are then executed on the GPU. The parameter refinement was then performed on the CPU, since \ac{BFGS} is inherently a sequential operation. Keeping it unchanged from the original implementation allowed for the hybrid architecture to be evaluated in isolation. The hybrid pipeline is shown in Fig.~\ref{fig:act_hybrid_pipeline}.

This work retains CuPy as the GPU backend from prior work and additionally adds a PyTorch backend. CuPy provides a NumPy-compatible interface on NVIDIA GPUs, offering minimal overhead and avoiding repeated transfers. PyTorch was included for downstream machine learning pipelines, which usually expect inputs in the form of torch tensor objects, and to enable cross-platform deployment via the ROCm backend for AMD and the Metal backend for Apple Silicon. Both AMD validation and Apple Silicon were attempted. However, the available older AMD GPU provided limited ROCm support for matrix operations required for the chirplet projection and required a Linux operating system, and the Metal backend currently lacked support for complex-valued operations. Therefore, both are deferred to future work as backend and hardware acceleration support matures. 

\subsection{Methodology}
The Bitbrain EEG dataset and the preprocessing pipeline described in Section~\ref{sect:Algorithm} were used. This included the notch filter at 50 Hz and the bandpass filter between 0.1 and 20 Hz. The \ac{ACT} parameter ranges were also kept the same as in Table~\ref{tab:parameters_clean} in Section~\ref{sect:Algorithm}; the only exception was the step size of the time centre, which was increased from 16 to 32. This was to reduce memory overhead, allowing both the primary results for the 3-second (768 samples) and 1-second (256 samples) epochs to be included. The 3-second epoch was used to better illustrate the difference between the two implementations in tabular form. However, the 1-second results were included to show that the findings were not specific to that window length. These epochs were all done with an order of 10, meaning 10 chirplets were used to represent the signal. 51 consecutive, non-overlapping windows were used, with the first epoch serving as a warm-up for GPU initialization and excluded from the resulting metrics. This warmup period occurs so that the CPU can establish a connection with the GPU, load any drivers, enable Just-In-Time compilation, and ramp up the GPU to the high-performance power state. The remaining 50 epochs were evaluated across 5 runs. The experiment was conducted on a desktop system with an NVIDIA RTX 5070 GPU, an i9-14900K processor, and 64 GB of system memory.

\subsection{Results and Discussion}
\begin{figure}[!ht]
\centering
\includegraphics[width=\linewidth]{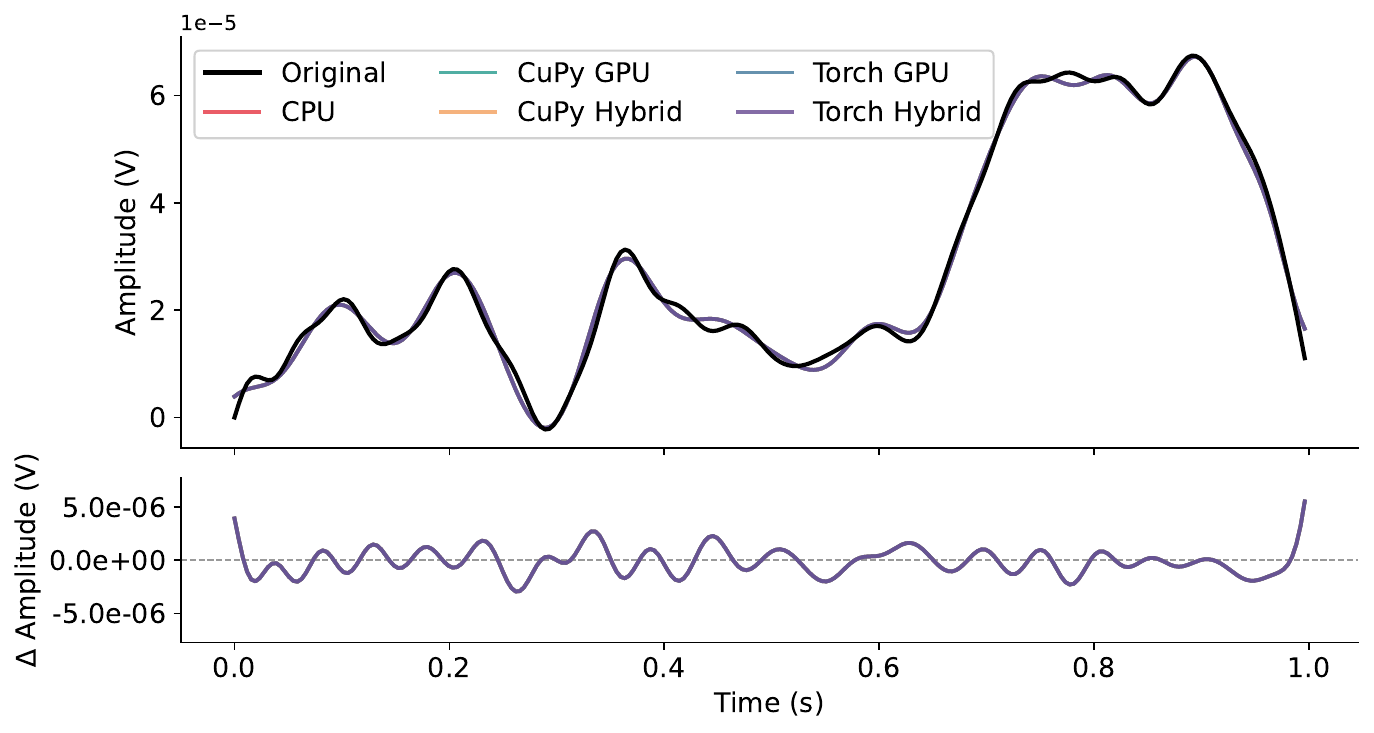}
\caption{Visualization of reconstructed signals across CPU, GPU, and hybrid implementations, confirming numerical equivalence across all modes.}
\label{fig:model_comparison}
\end{figure}

Before examining the speedup of the hybrid architecture, it is extremely important to first verify numerical equivalence across the implementations. This is confirmed in Figure~\ref{fig:model_comparison}; the upper panel shows that all five backends reconstruct the EEG signal with results visually indistinguishable from those of the CPU implementation presented in Section~\ref{sect:Algorithm}. The lower panel shows the differences between the four backends and the CPU implementation, again showing no visual difference. Across all 50 epochs, the normalized residue for the pure-PyTorch implementation yields $0.1245\pm0.0239$, and the PyTorch hybrid implementation yields $0.1256\pm0.0236$, whereas both CuPy implementations and the CPU implementation yielded a normalized residue of $0.1255\pm 0.0236$. These minute differences can be attributed to floating-point arithmetic differences across backends rather than any algorithmic differences. This confirms numerical equivalence in practice. 

After verifying that accuracy is not compromised by modifying the tensors or arrays, it becomes vital to examine runtime performance. Table~\ref{tab:benchmark} summarises the benchmark results at three-second epochs. Both hybrid implementations achieve the best overall runtime performance, $8.372 \pm 0.120$s and $7.952 \pm 0.016$s for CuPy Hybrid and Torch Hybrid, respectively. Compared with $15.342 \pm 0.238$s for the CPU baseline, there is an overall speedup factor of approximately $1.9\times$. Pure GPU implementations achieve shorter epoch times but longer overall times, consistent with previous results~\cite{11512436}. It highlights an important finding: naively offloading an entire pipeline to the GPU will not always yield better results and can in fact degrade performance. In this case, the independent generation of chirplet maps is very poorly suited to the GPU, which the hybrid architecture can resolve by assigning the search stage to the GPU and keeping the generation on the CPU. 

For the one-second epochs, the hybrid for CuPy and PyTorch achieve per-epoch times of $0.009\pm0.0003$s and $0.007\pm0.003$s, respectively, compared to $0.019\pm0.002$s for the CPU baseline. The overall runtime of these hybrid methods is $1.743\pm0.015$s and $1.615\pm0.007$s, respectively, compared to $2.274\pm0.04$s for the baseline. This shows a more modest improvement, consistent with the expectation that the projection stage contributes more as signal length increases, meaning that the matrix $P = DX^T$ grows with $L$. The results confirm that the hybrid architecture makes ACT decomposition real-time and feasible in practice, without sacrificing numerical accuracy. To contextualize these results practically, consider that a standard clinical sleep EEG recording lasts eight hours, which can be represented as 28,800 one-second epochs. It can be assumed that the generation of chirplets occurs within the same time frame, since they both run on the CPU; the differentiating factor is the epoch time. With the CPU epoch time, processing would take approximately 8.5 minutes per channel, whereas the hybrid implementations reduce this to under 4.5 minutes. However, consumer EEG systems can operate from two to four simultaneous channels, and clinical systems can routinely operate from 24 to 128 simultaneous channels. Since this cost is incurred by each individual channel, the cumulative runtime can become increasingly significant as the channel count grows, as examined in the next section.


\begin{table}[!ht]
\centering
\resizebox{\linewidth}{!}{
\begin{tabular}{lccc}
\toprule
\textbf{Implementation} & \textbf{Epoch Time (s)} & \textbf{Overall Runtime (s)} & \textbf{Normalised Residue} \\
\midrule
CPU           & $0.166 \pm 0.023$ & $15.342 \pm 0.238$ & $0.1255 \pm 0.0236$ \\
CuPy GPU      & $0.025 \pm 0.005$ & $63.374 \pm 0.431$ & $0.1255 \pm 0.0236$ \\
CuPy Hybrid   & $0.025 \pm 0.005$ & $8.372 \pm 0.120$  & $0.1255 \pm 0.0236$ \\
Torch GPU     & $0.029 \pm 0.016$ & $29.419 \pm 0.300$ & $0.1245 \pm 0.0239$ \\
Torch Hybrid  & $0.022 \pm 0.005$ & $7.952 \pm 0.016$  & $0.1256 \pm 0.0236$ \\
\bottomrule
\end{tabular}
}
\vspace{1mm}
\caption{Benchmark results across five runs for three-second epochs. Overall Runtime refers to both the generation of the chirplet family and the search epochs for each run. Epoch time, full runtime and normalized residue are reported as mean $\pm$ std.}
\label{tab:benchmark}
\end{table}

Before moving on to the next section, it is worth discussing a method for using higher-resolution parameter grids. As the number of chirplets grows, it can approach or exceed the available VRAM. Under the allocator in CuPy or PyTorch, it produces a hard out-of-memory error that can terminate execution on local systems that rarely exceed 32 GB. Inspired by the Apple Silicon architecture, where the CPU and GPU share a single physical memory pool, CUDA unified memory was implemented. This does not replicate Apple Silicon's hardware architecture, but it provides an analogous software-level abstraction that allows the GPU to page memory from both VRAM and system RAM as needed. This allows local systems to execute even when the number of chirplets exceeds the available GPU memory. This does have a trade-off: increased per-epoch latency when accesses miss VRAM and must be resolved from system RAM. However, this provides a practical way to run higher-resolution parameter grids on local hardware without requiring additional high-capacity GPUs, and it can serve as an operational fallback when the number of chirplets cannot be reduced. 

\section{Multi-Channel Scalability}
The previous section demonstrated that the hybrid architecture can reduce processing time compared with the CPU baseline presented in Section \ref{sect:Algorithm}. As mentioned earlier, in practice, EEG systems can routinely operate within tens to hundreds of simultaneous channels. Therefore, to make this algorithm practically feasible, the decomposition must scale efficiently. This section examines whether the framework degrades under multi-channel load and whether any speedup occurs as the number of channels processed simultaneously increases. Since channel scaling is being addressed, Bitbrain's dataset introduced in Section~\ref{sect:Algorithm} cannot be used, as it is limited to two channels, which is insufficient to determine any scaling behaviour. Instead, the Children’s Hospital Boston, Massachusetts Institute of Technology (CHB-MIT) scalp EEG dataset~\cite{PhysioNet-chbmit-1.0.0} was used, as it provides a sufficient number of channels to evaluate scaling on a representative range of channels. 

\subsection{Methodology}
For subsequent experiments, the CuPy hybrid model was selected because the normalized residue aligned exactly with the CPU baseline across both epoch lengths, whereas the PyTorch hybrid exhibited a marginal deviation, which is undesirable for benchmarking and for achieving consistent, reproducible work. To enable simultaneous multichannel processing, the signals from each channel were stacked into a batched matrix, so that the projection can be processed in a single GPU operation. To serve as a fair reference against GPU batching, the CPU baseline processes each channel sequentially, consistent with the implementation in Section~\ref{sect:Algorithm}. In processing the data, the EEG signal was first filtered using a notch at 60 Hz to account for the North American power-line frequency and a bandpass filter between 1 Hz and 50 Hz. The ACT parameter ranges were kept the same as in Section~\ref{sect:accel}, except that the frequency centre spans 0.5 Hz to 45 Hz to capture the broader frequency content of this dataset, and, as in the Section, the time centre step size remains 32 samples. The epochs were done with an order of 10. A representative subset of 1, 2, 3, 4, and 10 channels was evaluated. 51 consecutive non-overlapping epochs were used, with the first epoch serving as a warm-up and excluded from the results. The remaining 50 epochs were evaluated across 5 runs. This was run on the same desktop presented in Section~\ref{sect:accel}.

\subsection{Results and Discussion}

Table~\ref{tab:channel_scaling} presents the scaling results across channel counts. As expected, the GPU speedup increases monotonically with channel count from $3.94\times$ at a single channel to $8.22\times$ at 10 channels. As mentioned earlier, stacking the input matrix amortizes the fixed kernel launch overhead across a large workload, thereby improving hardware utilization. The CPU time remains constant at $\sim0.066$s regardless of the channel count, indicating that processing is performed sequentially per channel. Looking at the 10 channels, the per-epoch runtime is slower on the GPU than the CPU per-epoch time since it processes all channels in a single batched operation, whereas the CPU processes all 10 channels in $\sim0.66$s due to scaling linearly, giving the overall GPU runtime of $4.174 \pm 0.019$s compared to $34.306 \pm 0.676$s on the CPU. Additionally, the normalized residue remains consistent across CPU and GPU for each channel count. This confirms that batched multichannel processing does not incur any issues with decompositions. The higher residual values simply reflect the different signal characteristics of seizure EEG in the CHB-MIT dataset compared to the Bitbrain dataset. This is not due to any methodological differences.  

To recontextualize this, consider the sleep EEG example from the previous section. Assume the EEG has 10 channels and uses the same ACT parameters presented in this section. Processing the eight-hour session, the CPU would take approximately 5.3 hours, and the hybrid with the multichannel reduces this to 39 minutes. As higher-density clinical EEGs can support up to 64-128 channels, this architecture is useful for practical deployment. The results of this hybrid multichannel architecture demonstrate a GPU speedup that scales with problem size, making it a practical foundation for high-channel decomposition pipelines.

\begin{table}[!ht]
\centering
\resizebox{\linewidth}{!}{
\begin{tabular}{cccccc}
\toprule
Device & Channels & All Runtime (s) & Epoch Runtime (s) & Norm Residue & Speed-up \\
\midrule
CPU & 1  & $3.391 \pm 0.165$ & $0.0664 \pm 0.0096$ & $0.325 \pm 0.133$ & --- \\
GPU & 1  & $0.860 \pm 0.176$ & $0.0151 \pm 0.0031$ & $0.325 \pm 0.133$ & $3.94$ \\
\midrule
CPU & 2  & $6.532 \pm 0.155$ & $0.0640 \pm 0.0078$ & $0.320 \pm 0.125$ & --- \\
GPU & 2  & $1.168 \pm 0.010$ & $0.0227 \pm 0.0041$ & $0.320 \pm 0.111$ & $5.59$ \\
\midrule
CPU & 3  & $10.064 \pm 0.191$ & $0.0658 \pm 0.0107$ & $0.337 \pm 0.115$ & --- \\
GPU & 3  & $1.521 \pm 0.008$ & $0.0296 \pm 0.0045$ & $0.337 \pm 0.091$ & $6.61$ \\
\midrule
CPU & 4  & $13.537 \pm 0.318$ & $0.0664 \pm 0.0103$ & $0.327 \pm 0.105$ & --- \\
GPU & 4  & $1.948 \pm 0.017$ & $0.0380 \pm 0.0047$ & $0.327 \pm 0.073$ & $6.95$ \\
\midrule
CPU & 10 & $34.306 \pm 0.676$ & $0.0673 \pm 0.0109$ & $0.314 \pm 0.099$ & --- \\
GPU & 10 & $4.174 \pm 0.019$ & $0.0811 \pm 0.0057$ & $0.314 \pm 0.065$ & $8.22$ \\
\bottomrule
\end{tabular}
}
\vspace{1mm}
\caption{Channel scaling: CPU vs. GPU runtime for different numbers of channels processed simultaneously. Speedup is relative to the corresponding CPU runtime at each channel count.}
\label{tab:channel_scaling}
\end{table}

\section{Hardware Scalability}
\label{sect:HardwareScalability}
Given that this hybrid architecture can run on high-end desktop workstations, the question arises of how realistic it is for this framework to run across a variety of hardware environments, such as consumer laptops and embedded edge devices. This section evaluates whether this architecture's speedup generalizes across this hardware spectrum, testing three representative platforms: the reference desktop used in previous sections, as described in Section~\ref{sect:accel}, with the RTX 5070; a consumer laptop equipped with an RTX 5060; and an NVIDIA Jetson AGX Xavier (2018). 

\subsection{Methodology}
For the data, the Bitbrain sleep EEG data is used. The preprocessing pipeline was identical to that in Sections~\ref {sect:Algorithm} and~\ref{sect:accel}, with the notch filter at 50 Hz and the bandpass filter between 0.1 Hz and 20 Hz. A single channel was analyzed using the ACT parameter ranges from Table~\ref{tab:parameters_clean}, with the only difference being the use of three-second epochs (768 samples) instead of 1-second epochs. Similarly to other sections, 51 consecutive, non-overlapping epochs per repeat were used. The first was excluded to account for cache warm-up. The remaining 50 epochs were evaluated across 5 runs. An additional step added for this specific use case, especially for the AGX Xavier, was to run three more warm-up epochs due to dynamic voltage and frequency scaling issues that occur when ramping up to their sustained performance state. The laptop system was equipped with an AMD Ryzen AI 9 (HX 370) processor, an NVIDIA RTX 5060, and 32 GB of system RAM. The Jetson AGX Xavier featured an NVIDIA Carmel ARMv8.2 CPU and an integrated Volta GPU with 32 GB of unified memory. 

\subsection{Results and Discussion}
Table~\ref{tab:hardware} details the runtime comparison on all three platforms. The high-end desktop (RTX 5070) showcases the highest speedup at $\times 7.38$, reducing the per-epoch time from $298.33 \pm 24.05$ ms to $40.43 \pm 3.66$ ms. This differs slightly from the early result, but this could be due to run-to-run variability across independent experiments. The laptop achieves a $\times 4.05$ speedup, demonstrating that meaningful acceleration is possible on laptop hardware. Interestingly, the CPU baseline epoch times were nearly identical between the two platforms. This confirms that any difference in speedup is attributable to GPU capability. 

On the other hand, the Jetson provides a completely different profile. The CPU baseline is extremely slow, at about twice the per-epoch runtime on the desktop and laptop. This is consistent with the 8-core ARM Carmel architecture's clock speed, which ranges from 1.4 to 2.2 GHz, whereas the i9 can reach up to 6 GHz~\cite{intel_i9_14900ks_specs, techpowerup_jetson_xavier}. The hybrid architecture achieves a modest $\times 1.41$ speedup. There are a couple of interesting ideas to dissect within this. Firstly, this indicates that the GPU and CPU are significantly less powerful than those on other platforms. Secondly, there is a higher standard deviation ($\pm 86.44$ ms) than either of the other platforms, indicating significant memory bus contention, as the CPU and GPU must share Xavier's limited 137 GB/second unified memory bandwidth, thereby starving resources in hybrid execution~\cite{techpowerup_jetson_xavier}. That being said, the hybrid architecture still yields a consistent speedup even on this constrained embedded platform. If newer Jetson platforms were used, it should dramatically improve acceleration factors and reduce variance, primarily due to the increased bus width, higher max bandwidth, and significantly more CUDA cores. 

\begin{table}[!ht]
\centering
\resizebox{\linewidth}{!}{
\begin{tabular}{llccc}
\toprule
\textbf{Device} & \textbf{Backend} & \textbf{Epoch (s)}  & \textbf{Speed-up} \\
\midrule
\multirow{2}{*}{RTX 5070 Desktop}
 & CPU          & $0.298 \pm 0.024$ & --- \\
 & CuPy Hybrid  & $0.040 \pm 0.004$  & $\times 7.38$ \\
\midrule
\multirow{2}{*}{RTX 5060 Laptop}
 & CPU          & $0.300 \pm 0.015$ & --- \\
 & CuPy Hybrid  & $0.074 \pm 0.006$  & $\times 4.05$ \\
 \midrule
\multirow{2}{*}{Jetson AGX Xavier (2018)}
 & CPU          & $0.544 \pm 0.102$ & --- \\
 & CuPy Hybrid  & $0.386 \pm 0.086$  & $\times 1.41$ \\
\bottomrule
\end{tabular}
}
\caption{Runtime comparison of CPU and CuPy hybrid implementations across multiple hardware platforms, showing total execution time and achieved speed-up relative to CPU baselines. Generation time is excluded as it is CPU-bound and hardware-independent.}
\label{tab:hardware}
\end{table}
These results show that even with caveats, the hybrid architecture can generalize over a representative spectrum of hardware. It definitely scales with GPU capability, as expected, but has remained positive across all devices. This section shows that this framework can be deployed in practice outside a controlled environment. 

\section{LEM Hierarchical Search}
\subsection{Algorithm}
As noted in the previous sections, one of the major issues was memory usage. Since each chirplet atom is stored as a full-length signal vector, it can easily exceed available VRAM as you increase resolution on parameter grids. To address this, a hierarchical coarse-to-fine search strategy, inspired by \ac{LEM}, is introduced as a preliminary exploration of memory-scalable search. Since the original algorithm would generate a large, dense parameter grid for the search to traverse, the goal of this algorithm was to produce small, tunable parameter grids around areas of strong signal correlation. 

\begin{figure}[!ht]
\centering
\begin{tikzpicture}[node distance=1.5cm, auto, font=\small, >=stealth]
\tikzstyle{block} = [rectangle, draw, fill=green!10, minimum width=3.5cm, 
    minimum height=1cm, align=center]
\tikzstyle{subblock} = [rectangle, draw, fill=gray!30, minimum width=3.5cm, 
    minimum height=0.8cm, align=center]
\tikzstyle{arrow} = [->, thick]
\node [block] (signal) {Input Residual Signal $r_k$};
\node [block, below of=signal] (lhs) {Coarse Sampling (LHS)\\ 
    Generate Initial Candidate Parameters};
\node [block, below of=lhs] (proj) {Compute Projections onto Residual};
\node [block, below of=proj] (rerank) {Rerank the Top Candidates Globally\\ 
    Retain Top-$k$ Candidates};
\node [block, below of=rerank] (refine) {Refinement Stage\\ 
    Local Neighborhood Dictionaries\\ Compute Projections};
\node [block, below of=refine] (grad) {Gradient-Based Optimization (BFGS)};
\node [subblock, below of=grad] (final) {Select Refined Chirplet Parameters\\ 
    Output Best Atom};
\draw [arrow] (signal) -- (lhs);
\draw [arrow] (lhs) -- (proj);
\draw [arrow] (proj) -- (rerank);
\draw [arrow] (rerank) -- (refine);
\draw [arrow] (refine) -- (grad);
\draw [arrow] (grad) -- (final);
\end{tikzpicture}
\caption{Hierarchical LEM-inspired coarse-to-fine search. The procedure begins with global LHS sampling, followed by reranking the global parameters, then local refinement and gradient-based optimization to select the final chirplet parameters.}\label{fig:lem_hierarchical}
\end{figure}
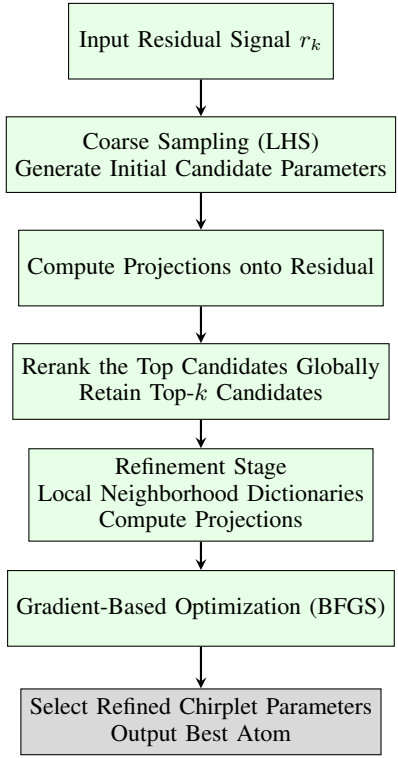

So imagine the four-dimensional space of the parameter grid; to find the best place to start, it uses structured random sampling via \ac{LHS}. This generates candidate chirplets, and each is projected onto the current residual; the best candidates (\textit{k}), based on the highest projection magnitude, are retained. In the subsequent refinement stages, a local chirplet family is constructed around the kept candidates. These bounds are user-defined, and the step sizes between parameters are usually finer as the refinement levels progress. Candidates from all parent neighbourhoods are then pooled together to globally rank the projection magnitude, and the best candidates are carried to the next level. Once again, all these values are user-defined, so it can be adapted to the signal and memory constraints. After the hierarchical refinement, the final candidate initializes \ac{BFGS} optimization. This pipeline is illustrated in Figure~\ref{fig:lem_hierarchical}.

\subsection{Methodology}
This section uses the same Bitbrain EEG dataset and the same preprocessing pipeline seen in other tables. This includes the notch filter at 50 Hz and the bandpass filter between 0.1 Hz and 20 Hz. Other than modifying the epoch length to 2 seconds (512 samples), the same parameter ranges were used as in Table~\ref{tab:parameters_clean}. The \ac{LEM}-inspired search was user-configured with 2 refinement levels, a top-$k$ of 4 candidates retained at each stage and a three level step size schedule of $(64, 2.0, 2.0, 5.0)$, $(32, 1.0, 1.0, 2.5)$, and $(16, 0.5, 0.5, 0.5)$ for the time-centre, frequency-centre, log-duration, and chirp-rate parameters, respectively. These values were chosen to enable an accurate comparison with the full-grid architecture. However, the step sizes could easily be reduced without any memory issues, whereas the previous full-grid hybrid architecture would encounter issues and exceed available VRAM at the equivalent resolution. 

Furthermore, the radius steps chosen were 2, meaning that the local search window extended in two coarse steps in each direction for all parameters other than log duration, which was one step in every direction. These values were selected to demonstrate a representative memory--accuracy--speed trade-off rather than optimized hyperparameters. It serves as a baseline configuration that can be refined for specific applications. Peak memory was monitored via a background thread that was sampling CPU resident set size and GPU memory pool usage at 10 ms intervals throughout each transform. 20 epochs were run, with the first excluded; the remaining 19 were evaluated repeatedly 5 times. A lower number was chosen due to the overhead of the memory-monitoring thread. These implementations were all benchmarked on the same desktop system described in Section~\ref{sect:accel}.

\subsection{Results and Discussion}
\begin{figure}[!ht]
\centering
\includegraphics[width=\linewidth]{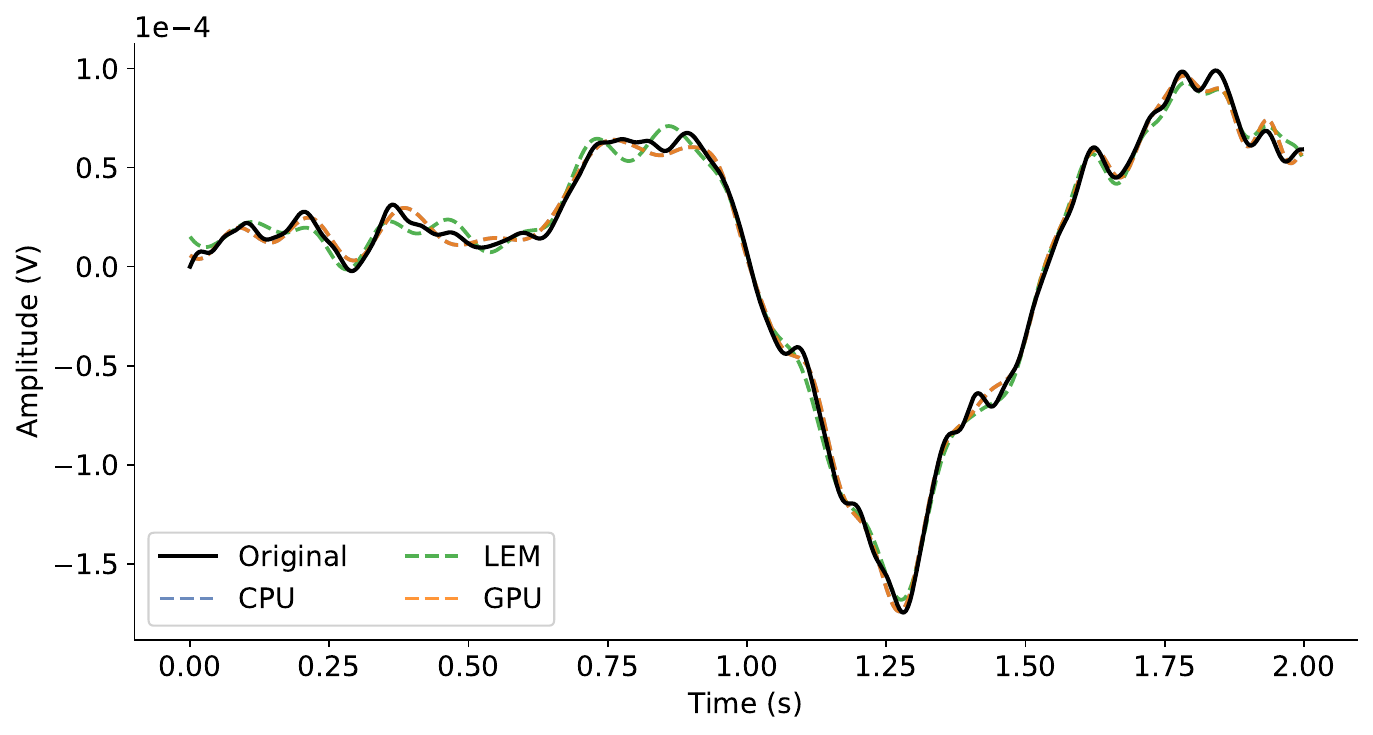}
\caption{Comparison of hierarchical LEM-inspired ACT and dense GPU implementations. The hierarchical method drastically reduces memory usage while producing EEG signal reconstructions closely matching the dense GPU results.}
\label{fig:lem_cpu_gpu}
\end{figure}

Table~\ref{tab:lem_tradeoffs} and Figure~\ref{fig:lem_cpu_gpu} summarise the results of the \ac{LEM}-inspired hierarchical search relative to the CPU and GPU baselines. As expected, the hierarchical search peaks at 999.4 MB of total memory (includes system memory and VRAM). In contrast, CPU required 2.62 GB, and the GPU required 4.76 GB to run the same process. The reduction in memory for hierarchical search is approximately $2.5\times$ less than CPU and approximately $4.5\times$ less than the GPU. Furthermore, it reduces the memory requirement to under 1 GB, which is viable on hardware with very limited VRAM or system memory. 

In addition, reconstruction quality remains competitive, with hierarchical search achieving a normalized residual of $0.094 \pm 0.026$ relative to $0.079 \pm 0.021$ for both CPU and GPU baselines. Looking at Figure~\ref{fig:lem_cpu_gpu}, the hierarchical search closely tracks the original signal, with small deviations during rapid amplitude changes, which is slightly distinct from the full-grid version. Upon closer investigation, the normalized residue had run-to-run variability. Both of these behaviours were expected. The small deviations arise because restricting the candidates to sampled regions may provide a locally optimal chirplet rather than the globally optimal one. The run-to-run variability can be mechanistically traced to \ac{LHS}, where the initial candidates can be completely different, thereby propagating to different locally optimal chirplets. 

\begin{table}[!ht]
\centering
\resizebox{\linewidth}{!}{
\begin{tabular}{cccc}
\toprule
System & Epoch (s) & Total Memory (MB) & Norm Residue \\
\midrule
LEM & $11.46 \pm 0.84$ & $999.4$  & $0.094 \pm 0.026$ \\
CPU & $0.147 \pm 0.013$ & $2623.0$ & $0.079 \pm 0.021$ \\
GPU & $0.023 \pm 0.003$ & $4755.6$ & $0.079 \pm 0.021$ \\
\bottomrule
\end{tabular}
}
\caption{Comparison of LEM, CPU, and GPU performance with 2-second epochs. Values are mean $\pm$ std.}
\label{tab:lem_tradeoffs}
\end{table}

Even with the benefits of this technique of being below 1 GB, there is a very large tradeoff, and that is speed. Looking at Table~\ref{tab:lem_tradeoffs}, the mean epoch runtime is approximately $11.46\pm0.84$, which is substantially higher than the CPU and GPU baselines, at $ 0.147\pm0.013$ and $0.023\pm0.003$, respectively. This is just the search component, so for every two seconds it takes approximately eleven seconds to process the data; this reflects the multi-stage candidate evaluation, the construction of the small family of chirplets, and each refinement level. It introduces sequential dependencies which limit the usage of GPU parallelism. This is a tunable system, of course, so there are ways to achieve more accurate decompositions, require less memory, or take less time, allowing individuals to adjust to their constraints. 

Once again, this is a section for preliminary exploration of memory-scalable search for \ac{ACT}, rather than a fully optimized version. There is still a lot of work left on parameter scheduling, candidate-pruning strategies, and the construction of local families of chirplets to make this framework as practical as the accelerated hybrid framework. Nevertheless, these results demonstrate a way to decompose signals using the \ac{ACT} that can be achieved with a memory-constrained system. 

\section{Discussion}
\label{sect:Discussion}
\begin{table*}[t]
\centering
\resizebox{\linewidth}{!}{
\begin{tabularx}{\textwidth}{cc>{\centering\arraybackslash}X>
{\centering\arraybackslash}X>{\centering\arraybackslash}X>
{\centering\arraybackslash}X>{\centering\arraybackslash}X}
\toprule
Contribution & Issue Addressed & Impact & Evidence\\
\midrule
Unit Normalization and Residual Projection & MP theory alignment & Lower Normalized Residue & Table~\ref{tab:performance_summary_clean}\\
GPU Acceleration & Compute Bottleneck & Per Epoch Speedup of 6.64 & Table~\ref{tab:benchmark}\\
Multi-Channel & Single Channel Limitation & Simultaneous Multi-Signal
Processing, $\times8.22$ at 10 Channels & Table~\ref{tab:channel_scaling}\\
Hardware Scalability & Device Dependence & Edge Device Deployment & Table~\ref{tab:hardware}\\
LEM Search & Memory Footprint & Sub-1 GB Operation, Reconstruction Tradeoff Remains & Table~\ref{tab:lem_tradeoffs}\\
\bottomrule
\end{tabularx}
}
\caption{Summary of contributions, issues addressed, observed impact, and supporting evidence across all experimental sections.}
\label{tab:summary}
\end{table*}

The contributions of this paper follow a deliberate structure, with each section addressing a bottleneck identified in the previous one. It had to start with the decomposition because, without an \ac{ACT} aligned with theory, nothing else would provide much value. The propagation of error would occur through the parallelization, multichannel stacking and hardware portability. The prior implementation was a source of structural instability that would have undermined any acceleration this framework could provide. To fix this issue, unit normalization and residual projection were used. This established a foundation that ensured the best chirplet could consistently be extracted from the signal. The results in Table~\ref{tab:performance_summary_clean} validate this statement with minimal domain-specific tuning. As an unintended consequence, the runtime did improve. This is most likely due to an implementation-level correction to remove redundant precision casting in the previous codebase, where it was initially set to float64 and then converted to a float32 array. 

After determining that stable decomposition could be achieved, the next goal was to extend it into a practically deployable framework. Building on previous work, it was determined that the CPU was better at generating chirplets and the GPU was better at performing search. When building this hybrid architecture, GPU parallelization reduced per-epoch time by a factor of 6.64 (and could reach 7.38, as seen in Section~\ref{sect:HardwareScalability}, consistent with run-to-run time variability), thereby avoiding the kernel launch overhead that slowed the pure-GPU generation overall. This enabled multichannel stacking, where the projection matrix can naturally extend to batch channels, processing multiple channels simultaneously. In practice, this can be valuable, especially when many clinical devices require multiple channels, and having a solution that scales acceleration with increasing channel count can make \ac{ACT} very useful for deployment. Regarding practical deployment, the previous sections make it clear that on a high-end desktop it can accelerate, but some specific scenarios may require different deployment environments where desktops are unavailable. Testing on the Jetson and the RTX 5060 still showed meaningful speedups.

Previous papers overlooked memory as an axis of scalability, usually treating it as a secondary concern relative to runtime. However, the goal of the \ac{LEM}-inspired section was to address this directly by reducing peak memory to $999.4$ MB while maintaining competitive reconstruction accuracy. However, as a trade-off for speed, a concrete decision framework needs to be developed across different deployment scenarios. When sufficient VRAM is available, the GPU hybrid is the best choice; however, if a GPU is present but memory is constrained, LEM can provide a viable path to decomposition. The speed cost may be substantial, but parameter modifications such as top-$k$, refinement levels, and step size can slightly accelerate runtime while potentially reducing reconstruction accuracy. This tunable system can improve reconstruction accuracy or reduce memory or runtime, based on the constraints of the deployed system. 

The differences in normalized residues across domains warrant discussion. Rather than a failure of the framework, this highlights differences in signal complexity across signals. In sleep \ac{EEG}, it is relatively stationary and well-structured, which means that at low decomposition order and a more compact family of chirplets, it can be decomposed very well, whereas for micro-Doppler signatures or \ac{EMG}, they may exhibit more complex behaviour which requires finer parameter grids or higher decomposition orders. This is consistent with Figure~\ref{fig:ordervsresidue}, where each of the signals' normalized residue decreases as more decomposition orders are added. This motivates the choice of parameter ranges used in this work, which were selected as conservative, reproducible defaults rather than optimized values. This was deliberate so that the framework can remain domain-agnostic, while still showing meaningful decomposition across a variety of signals without fine-tuning. It serves as a reproducible baseline where domain experts can refine it to more accurately reflect the specific characteristics of their signal or application. Ultimately, this framework can perform consistently across all domains, providing a starting point for a broad class of non-stationary signals. 

Finally, it becomes important to be precise about what this paper does and does not claim. The contributions of this paper are correctness, stability and acceleration of the \ac{ACT} framework across multiple signal domains and hardware environments. It does not claim that this parameter range is optimal for any specific domain. The optimality of the parameter grid for a specific task is an application-specific question that depends on the signal's spectral and temporal characteristics. The framework introduced in this paper is stable, scalable and accessible across a wide range of hardware and constraints.

\section{Conclusion}
\label{sect:Conclusion}
This paper presented a correct, stable and practically deployable \ac{ACT} framework based on a series of contributions. First, it establishes a foundation of correctness through unit normalization and residual projection. With the behaviour finally consistent with \ac{MP} theory, a hybrid CPU--GPU architecture was implemented to assign generation to the CPU and projection search to the GPU, achieving up to a $\times 7.38$ speedup over the CPU baseline on a desktop RTX 5070. This was then extended to multichannel batching. The per-epoch runtime speedup increased from 3.94 for a single channel to 8.22 for 10 channels. The scalability of the acceleration was then tested on consumer laptops and a 2018 Jetson Xavier embedded system, validating that meaningful speedups could be achieved outside of high-end desktops. Finally, the \ac{LEM}-inspired hierarchical coarse-to-fine search introduced a preliminary version of a memory-scalable decomposition, reducing peak memory usage to below 1 GB at the expense of increased runtime while maintaining reconstruction quality comparable to dense parameter-grid baselines. 

There are limitations to the current work, but these point to concrete directions for future investigation. Currently, the \ac{BFGS} refinement step runs on the CPU. This means it requires an extra device transfer at each decomposition order. It may make minimal difference on high-end desktops, but with something like the Jetson Xavier, where memory bandwidth is scarce, removing that bottleneck may be extremely useful. Replacing it with a GPU-native optimizer, such as a gradient descent variant, could work but would require much further research. As mentioned earlier, it is currently just a framework, so the next step is to work towards downstream applications that were previously impractical at CPU speeds. Currently, seizure prediction from scalp \ac{EEG} is a planned application under investigation. 

At the implementation level, the current implementation relies on CuPy's memory abstraction, which provides a practical high-level interface. However, implementing lower-level \ac{CUDA} kernels to control unified memory or memory transfers can reduce host-device communication overhead. This is once again looking at edge devices such as the Jetson, where memory transfer was a bottleneck in runtime performance. Furthermore, validation of the PyTorch backend on both AMD and Apple Silicon hardware is still necessary to improve accessibility.

The \ac{LEM}-inspired hierarchical search also offers several directions for optimization, especially since it has been a preliminary exploration so far. Strategies to improve accuracy could include simulated annealing. It may yield higher-quality candidates than a greedy local search. Switching to a GPU-based optimizer could also improve search speed and prevent memory transfers. Further work will have to explore vectorized parameter grids to reduce generation overhead.  Overall, this paper provides a foundation on which more targeted optimization and application of the \ac{ACT} can be built. 

\bibliographystyle{ieeetr}
\bibliography{IEEEfull}

\end{document}